\magnification= 1200
\def\rv{{\bf r}}

\def\Ham{{\cal H }}
\def\OP{{\cal O }}

\def\N{{\cal N }}
\def\K{{\cal K }}
\def\EF{{\cal E}}
\def\Trans{{\cal T }}

\def\Energy{{ E }}

\def\Flip{{\cal F}}
\def\bra{\left<}
\def\ket{\right>}

\def\der{{\partial}}
\def\bra{{\langle}}
\def\ket{{\rangle}}
\def\half{{\scriptstyle{1\over2}}}
\def\3half{{\scriptstyle{3\over2}}}
\def\quart{{\scriptstyle{1\over4}}}

\hsize= 6.0 true in
\vsize= 8.5 true in
\baselineskip 18 pt plus 0.2 pt minus 0.2 pt

\centerline{\bf STATIONARY STATE SKEWNESS IN KPZ TYPE GROWTH }
\bigskip
\centerline{\bf John Neergaard and Marcel den Nijs} 
\bigskip
\centerline{\it Department of Physics} 
\centerline{\it University of Washington} 
\centerline{\it Seattle, WA 98195}
\vskip 50 pt 
\noindent
\baselineskip 15 pt plus 0.2 pt minus 0.2 pt

Stationary states in KPZ type growth have interesting
short distance properties.
We find that typically they are skewed and lack particle-hole symmetry. 
E.g., hill-tops are typically flatter than valley bottoms, and all
odd moments of the height distribution function are non-zero. 
Stationary state skewness can be turned on and off in the 1+1 
dimensional RSOS model.
We construct the exact stationary state for its master equation in a 
4 dimensional parameter space. 
In this state steps are completely uncorrelated.
Familiar models such as the Kim-Kosterlitz model lie
outside this space, and their stationary states are skewed.
We demonstrate using finite size scaling
that the skewness diverges with systems size, 
but such that the skewness operator is irrelevant in 1+1 dimensions,
with an exponent $y_{sk}\simeq-1$, and that the 
KPZ fixed point lies at zero-skewness.

\vfill
\eject
\baselineskip 18 pt plus 0.2 pt minus 0.2 pt

\centerline{\bf 1. INTRODUCTION}
\bigskip

Crystal surfaces display interesting scaling properties during growth. 
One of the dynamic processes that has been at the center of attention 
is so-called KPZ type growth, named after the Langevin equation
$$
{d h \over d t} = v_0 +\nu \nabla^2 h + {1\over2} \lambda (\nabla h)^2 + \eta
\eqno(1.1a)
$$
with uncorrelated Gaussian noise
$$
\left< \eta(\rv_1,t_1) \eta(\rv_2,t_2) \right> = 2D \delta(\rv_1-\rv_2)\delta(t_1-t_2)
\eqno(1.1b)
$$
studied by  Kardar, Parisi, and Zhang [1].
The  growth rate $v_0$ is modified by 
the local curvature of the surface (the $\nu$-term),
its local slope (the $\lambda$-term),
and random fluctuations (the $\eta$-term).
It is well established by now that many microscopic growth processes 
belong to the KPZ universality class [2-5].
Such microscopic models have been investigated by 
numerous Monte Carlo (MC) simulations and 
finite size scaling (FSS) studies using exact diagonalization of master equations.
Moreover, at least one model is exactly soluble in 1+1 dimensions (D=1) [6-9].
Intriguing relations with different aspects of physics have been established.
Eq.(1.1) is equivalent to the Burgers equation for randomly stirred fluids
($\nabla h$ represents the fluid velocity) [10].  
KPZ growth maps onto to the directed polymer problem [11,5],
and relates to equilibrium liquid crystal phases [12].
In 1D it describes  persistent currents in metal rings
in the context of asymmetric exclusion models [13-17], 
and is also equivalent to the equilibrium statistical mechanics 
at facet-ridge end-points in 2D crystals [13].

The scaling properties of KPZ type growth have been established,
but our understanding is not yet at the level we would like.
It is useful to make a comparison with equilibrium critical phenomena. 
Master equations for microscopic models, like the 
body-centered solid-on-solid (BCSOS) model 
and the restricted solid-on-solid (RSOS) model, 
play a similar role as Ising type microscopic models in  
equilibrium critical phenomena.
KPZ type Langevin equations play a similar role as 
$\phi^4$ type field theories.
There are roughly three levels at which a particular type of 
scaling invariance can be understood.
The first level is to establish empirically 
(experimentally and numerically)
the existence of scale invariance and universality. 
For equilibrium critical phenomena this was achieved in the early seventies,  
for KPZ type growth only a few years ago.

The second level is to obtain analytical confirmation of 
the empirical scaling properties. 
Exactly soluble models confirm the empirical scaling properties of 2D
equilibrium critical phenomena.
Mean field approximations and Landau-Ginzburg theory yield 
the existence of an upper critical dimension 
and makes it possible to carry out controlled renormalization 
transformations that demonstrate the 
existence of the Ising fixed-point in $\phi^4$ theory.
For KPZ type growth, we only reached this level in 1D.
The BCSOS growth model is exactly soluble in 1D.
The KPZ fixed point in general D has eluded us thus far. 
The existence of an upper critical dimension is yet unclear.
The $\epsilon$-expansion analysis of eq.(1.1) [1,10,18] describes the 
reversal of the stability of Edwards-Wilkinson (EW) growth
(the point $\lambda=0$)  with respect to $\lambda$ at $D=2$.
It does not yield a fixed point for the KPZ universality class, except 
a strong coupling one in 1D.

The third level is to express the scaling properties
in terms of a free field theory.
In general this is not possible at all.
However, for 2D equilibrium critical phenomena such a description emerged 
during the last 15 years.
Coulomb gas representations [19,20] and 
conformal field theory [21] provide a full free field theory
description of virtually all 2D equilibrium phase transitions.
There is no assurance that this
can be generalized  to dynamic processes in 1D,
but we have a good chance
since the time evolution operators of master equations 
in 1+1 dimensions resemble closely transfer matrices of 2D 
equilibrium critical phenomena.
Generalizations of conformal invariance are being considered [22,23],
but it is too early to tell whether this will work.

In this paper we present a master equation study of a generalized
RSOS growth model.
In this model, nearest neighbour columns of particles 
can only differ by $dh=0,\pm 1$. 
We consider the most general growth rule involving
only nearest neighbour step configurations. 
This gives rise to a 5 dimensional phase diagram.  
We address four issues:
(a) we describe the global structure of this phase diagram:
(b) we derive the exact analytic form of the stationary state in a 
4 dimensional subspace; 
(c) we point out that the stationary state is generically skewed; and 
(d) we investigate whether stationary state skewness scales in accordance 
with conventional renormalization theory and the concept of universality.

Somehow, stationary state skewness seems to 
have escaped everyone's attention thus far.
The only discussions of skewness in the literature of which we are aware, 
concern temporal skewness [24-26,5].
One possible reason is that almost all previous studies involve MC 
type simulations. In master equation studies, like ours, 
the stationary state and its properties are directly accessible 
(as the eigenvector of the largest eigenvalue 
of the time evolution operator). 
Consider the moments of the height distribution $h_x(t)$ at a certain 
moment in time,
in a finite 1D system with periodic boundary conditions, 
$h_{x+L}(t)=h_x(t)$,
$$
W_n(L,t) = L^{-1} \int dx[ h_x(t)-h_{av}(t)]^n.
\eqno(1.2)
$$
The first moment vanishes 
since it defines the average surface height  $h_{av}(t)$ .
The second moment is the conventional measure for the width of the interface.
The third moment characterizes the skewness.
In EW type growth  all odd moments vanish because of 
particle-hole symmetry at $\lambda=0$.
KPZ type growth lacks that symmetry. 
On a local level, skewness means that  
the curvature at a typical valley-bottom is  
larger in magnitude than at a typical hill-top (or the other way around). 
Fig.1 illustrates this.
It shows the deterministic time evolution of a sinusoid initial state, 
according to eq.(1.1) without noise for positive $\lambda$.

It is important to distinguish between
stationary state skewness and temporal skewness.
The skewness in Fig.1 for the deterministic KPZ equation
is a transient phenomenon.
This initial state decays to zero, 
to a stationary state with no skewness whatsoever.
Noise pumps sinusoid waves at random frequencies into the surface
(from the spatial Fourier transformed perspective) at all time scales. 
Each mode decays with the same sign for skewness.
So in  the presence of random noise
the stationary state tends to be rough and skewed.
This  picture is too simplistic. 
It ignores the non-linearly coupling between the modes.
Moreover, the stationary state of eq.(1.1) is known exactly in 1D.
It is Gaussian. All moments $W_n$ for $n>2$ vanish [5,10,27].

The BCSOS model is one of the simplest 
microscopic KPZ type growth rules [28], and it is exactly soluble in 1D [6-9].
Its stationary state is trivial as well.
It is the completely disordered state,
without any skewness (nor any other structure),
not only in the thermodynamic limit but also for all finite $L$.
The origin of this is a special symmetry of the time evolution operator.
We review this in section 2.
Throughout this paper we will compare our results for the RSOS model
with the properties of the BCOS model.

Our study of the RSOS model illustrates that the simplicity of the stationary
states  of eq.(1.1) and the BCSOS model is accidental.
Generically, stationary states are skewed.
Skewness can be turned on and off in the RSOS model.
In section 3, we derive the exact form of its stationary state 
in a 4 dimensional subspace.
This state is simple.
Skewness and all other correlations between 
steps in the surface are absent.
Steps are placed at random, with not even nearest neighbour
correlations between them. The only parameter is the step density.
Familiar special points, such as the Kim-Kosterlitz (KK) model [29],
and speical lines [30] lie outside this subspace. 
Their stationary states are non trivial.
Stationary state skewness is one aspect of this.

In section 4, we present a mean-field type derivation of the
deterministic part of the KPZ equation for the BCSOS and RSOS model.
This derivation  provides more insight into the 
general structure of the RSOS model phase diagram.
In particular, it identifies a 
second order parameter, the step density, 
and a typical time scale, $\tau_s$. 
The step density (re-)equilibrates locally at a time scale $\tau_s$.
The RSOS model is desribed by the KPZ equation
only at time resolutions less than $\tau_s$.
In section 5, we combine these results into 
a description of the global structure of the RSOS model phase diagram.
We describe also how skewness can be introduced and controlled within 
the Langevin equation, eq.(1.1).

Stationary skewness raises an important fundamental level-2 type issue.
One of the  basic premises of renormalization theory that all 
short-distance complexities are expressible in terms of irrelevant operators.
Stationary state skewness should be an example of this.
It is important to demonstrate this explicitly, since level-2 type
understanding of KPZ type growth is still rudimentary. 
According to the empirical scaling theory,
the growing interface displays scale invariance
at large length scales and long times.
It is invariant under the transformation:
$x^\prime=b x$, 
$h^\prime = b^\alpha h$, and 
$t^\prime= b^z t$.
The moments of the height distribution scale as
$$
W_n(L,t) = b^{n\alpha} W_n(b^{-1}L, b^{-z}t)
\eqno(1.3)
$$
$\alpha$ is the stationary state roughness exponent and 
$z$ the dynamic exponent.
Galilean invariance in the Burgers equation representation  of eq.(1.1) 
yields the identity  $\alpha+z=2$ [1-5,10].
In 1D the height variable scales with the exponent $\alpha=1/2$.
The width of the surface, $W_2$,  
diverges linearly with $L$ in the stationary state.
The step-step correlation function 
$$
G_s(r)= \bra \nabla h_{x+r} \nabla h_{x} \ket  \sim \exp(-r/\xi_s)
\eqno(1.4)
$$
has a finite correlation length, or at least decays fast enough,
such that going up and down along the surface becomes a random
walk at large length scales.
It is unlikely that KPZ type models with stationary state skewness 
belong to a different universality as those without it.
The numerical evidence for $z=1.5$ is very strong,
in particular at the KK point.
But, are they all described by the same fixed point?

The conventional KPZ fixed point lies at zero skewness.
It applies to eq.(1.1), to the BCSOS model, and also to the
RSOS model inside the non-skewed 4 dimensional subspace.
This fixed point describes skewed surfaces as well,
if we can demonstrate that the skewness crossover operator is irrelevant in 
the sense of  renormalization theory.
This does not mean that aspects like skewness vanish in the thermodynamic limit.
Skewness $W_3$ is allowed to diverge with system size,
but at a rate slower than its naive exponent $3\alpha$. 
In section 5 we show numerically, 
using a master equation finite size scaling analysis,
that the skewness in the RSOS model 
diverges in the thermodynamic limit, as $W_3\sim N^x$, with $x\simeq 0.5$.
This confirms that the KPZ fixed point lies at zero skewness.
The skewness crossover scaling exponent is equal to $y_{sk}\simeq-1$.

\vfill
\eject

\centerline {\bf 2. THE STATIONARY STATE IN THE BCSOS MODEL}
\bigskip
The stationary state of the BCSOS growth model is simple.
It the completely disordered state.
This is a consequence of the symmetry properties of the master equation.
First we review these symmetries and then 
rederive the stationary state by a more complex method, 
one that can be generalized to the RSOS model in section 3.
  
Consider a surface built from rectangular shaped bricks
(a conventional brick wall turned over $90^\circ$).
The surface heights at $x=2n+1/2$ are even integers and those at 
$x=2n-1/2$ are odd. 
Nearest neighbour columns differ in height by one unit.
Each bond $x=n$ contains a step $S_n=\pm1$.
The growth rule is as follows:
Choose one of the columns at random.
If this column is at the bottom of a local valley, $S_{n-1}=-1$ and $S_{n}=+1$, 
a particle adsorbs  with probability $p$
(and nothing happens with probability $1-p$).
If it is at the top of a local hill, $S_{n-1}=+1$ and $S_{n}=-1$, 
a particle desorbs with probability $q$
(and nothing happens with probability $1-q$).
If it is part of a local slope, $S_{n-1}=S_{n}$, nothing happens.
This model has been studied extensively in the literature.
First by Monte Carlo simulations [28,3]. 
More recently it was realized it can be solved exactly [6-9].

The Master equation 
$$
|\Psi\ket _{t+1} = \Trans |\Psi\ket _{t} 
\eqno(2.1)
$$
describes the time evolution of the probability distribution
$c(\{S_n\})$
$$
|\Psi\ket =  \sum_{\{S^z_n\}} c(\{S^z_n\})~|\{S^z_n\}\ket
\eqno(2.2)
$$
The time evolution operator has the familiar form
$$
\Trans = 1-  N^{-1}  \sum _n \Ham(n,n+1) 
\eqno(2.3)
$$
but $\Ham$ is not Hermitian.
$$
\eqalignno{
\Ham(n,n+1) = {1 \over 4} \epsilon
[ 1 - S^z_n S^z_{n+1} 
- 2   &(S^+_n S^-_{n+1} + S^-_n S^+_{n+1}) \cr
- 2 s &(S^+_n S^-_{n+1} - S^-_n S^+_{n+1})] & 
(2.4)\cr}
$$
with $s = (p-q)/(p+q)$, 
and periodic boundary conditions, $S^z_{n+N}=S^z_n$.
$\Trans$ must be applied $N$ times to evolve the system
by one  unit of time.
Without loss of generality we can set $\epsilon = p+q =1$.

At $s=0$,
$\Ham$ is identical to the Hamiltonian of the so-called
XXZ quantum spin-$\half$ chain.
For any value of $s$ the model is equivalent to
the 2D equilibrium 6-vertex model in an electric field $s$.
To be more precise, 
the time evolution operator is identical to the transfer matrix of the 
6-vertex model when the sites are being updated sequentially instead of at random.
That master equation reduces to eq.(2.3) in  the time continuum limit.
The 6-vertex model and  eq.(2.4) are exactly soluble [7,8,31-33].
The master equation follows a special line through the 6-vertex model phase diagram,
where $\Trans$ is a stochastic matrix
($\Trans$  preserves probability).
The 6-vertex model describes the temperature 
evolution of equilibrium crystal surfaces [32].
KPZ type growth maps onto facet-ridge endpoints [13, 31], 
special points in the phase diagram where 
the rough  (chiral-Luttinger liquid) and faceted phases meet.

The stationary state of eq.(2.3) is very simple.
It is the disordered state 
$| D\ket$ where all coefficients $c(\{S_n\}\ket$ are equal.
For non-growing surfaces, with $p=q$, this is obvious. 
The dynamic rule is then equivalent to a Monte Carlo process 
in equilibrium statistical mechanics.
The coefficients of the stationary state are  proportional to
the equilibrium Boltzmann weights.
$|D\ket$ is the equilibrium state 
since this 1D  BCSOS model lacks any interactions. 

$|D\ket$ is the stationary state for the growing BCSOS surface as well.
Stochastic processes preserve probability.
Algebraically, this is expressed by the property that 
$\bra D |$ is the left eigenvector of the largest eigenvalue $\lambda=1$
for all  stochastic time evolution operators.
In general, the adjoint of a time evolution operator is not stochastic,
but for the BCSOS model it is.
$\Trans^\dagger$ describes the same growth process, 
but with the role of particles and holes reversed.  
The left and right eigenvectors switch role, and therefore
$|D\ket$ is the right eigenvector for all $s\neq0$.
The stationary state is completely disordered.

The fact that $\Trans^\dagger$ is stochastic in the entire phase diagram
is an accident.
It is also an accident that $\Trans^\dagger$ is identical to the
particle-hole transformed time evolution operator.
We will see this clearly in section 3 during the discussion of the 
RSOS model.
For that model we will be able to generalize the following method to
derive the stationary state.

The BCSOS model can be interpreted as a model for (classical) particles 
hopping along a chain in an electric field.
$S^z_n=1$ represent an occupied site and $S^z_n=-1$ an empty site.
This is also  known as the asymmetric exclusion model [14].
Consider the probability $c(\{S_n\})$ to be in micro state $\{S_n\}$.
Let the dots in $c(..,S^z_n,S^z_{n+1},..)$ mean that all spins other
than those at sites $n$ and $n+1$ are the same for all states considered
and do not change.
The action of each individual $\Ham(n,n+1)$,
$$
c(..,+,-,..)_{t+1/N} ~=~(1-q)~c(..,+,-,..)_t~~ +~~ p~c(..,-,+,..)_t
\eqno(2.5)
$$
leaves $|\Psi\ket$ invariant when
$$
{c(..,+,-,..) \over c(..,-,+,..)} = {p\over q}
\eqno(2.6)
$$
The electrostatic energy 
of charged particles hopping along a wire in an electric field $\EF$
is equal to $\EF\K$, with 
$$
\K ~=~\sum_n \half n~(S^z_n+1) 
\eqno(2.7)
$$  
A conventional  MC simulation rule for such a system obeys detailed balance,
in which the transition probabilities 
between states $a$ and $b$ are related as
$$
{P_{a \to b} \over P_{b \to a}} 
~=~\exp~(\EF~\Delta\K/k_BT). 
\eqno(2.8)
$$
This is the precisely the content of eq.(2.6).
The stationary  state is 
the ``atmospheric law" density distribution:
$$
\rho (n)~=~ \exp(-\EF n/k_BT)
\eqno(2.9)
$$
and  $p/q~=~\exp~(\EF/k_BT)$.
This is not the solution we are looking for.
It applies to the wrong boundary conditions. 
Eq.(2.9) is valid for open boundary conditions.
In our case the particles keep running around 
in a circle and no density profile can build up to stop this flow.
The detailed balance approach is apparently too restrictive.
We must attack the problem less locally.

The following property will prove essential.
The number of hill tops and valley bottoms is the same in every configuration.
This is easily established by drawing a typical configuration
with periodic boundary conditions and a specific average slope.
Algebraically it can be shown as follows.
Consider the four  nearest neighbour step-step densities in each configuration: 
$d_{+-}$, $d_{-+}$, $d_{++}$, and  $d_{--}$.
They  represent respectively,
the density of hill tops, valley bottoms, and  up and down slopes.
The density of up-spins, $d_+$, is related  by normalization as
$$
d_+=  d_{++} +  d_{-+} = d_{++} +  d_{+-} 
\eqno(2.10)
$$
This implies that  $d_{+-}=d_{-+}$.

Let's test the following assumption:
The probability distribution $c(\{S_n\})$ depends only on 
the total number of particles $\N$ and the total electrostatic energy $\K$.
The average slope of the surface is fixed by the boundary conditions.
Therefore $\N$ is a constant of motion and can be ignored.
This leaves only the dependence on the electrostatic energy,
$c(\{S_n\})=c(\K)$.
This probability distribution time evolves according to eq.(2.3) as
$$
{\der c(\K)\over \der t} = 
-(p d_{-+} + q d_{+-}) c(\K) +p d_{+-} c(\K+1) + q d_{-+} c(\K-1) 
\eqno(2.11)
$$
The first term on the right hand side represents the
contributions for when nothing happens between $t \to t+1/N$.
The second and third term represent events where one particle adsorbs
or evaporates.
Adsorption destroys one valley bottom and creates  one hill top.
Evaporation does the opposite.
The adsorption probability is proportional to $d_{+-}$
since this could have happened to any of the  hill tops
in existence  at time $t+1/N$.
It is also proportional to $C(\K+1)$, because
in the electric field interpretation
one particle hops in the direction of the electric field.

Our assumption for the stationary state is correct if
the right hand side of eq.(2.11) vanishes for all $d_{i,j}$.
This seems to imply two conditions, 
but there is only one, since
$d_{-+}$ can be eliminated using the identity $d_{-+}=d_{+-}$.
The right hand side vanishes for all configurations if
$$
p+q= p/w +qw
\eqno (2.12)
$$ 
with $w = c(\K)/c(\K+1)$.
This yields two stationary state solutions:
the detailed balance solution with $w=p/q$, and 
the stationary growing state solution $|D\ket$ with $w=1$,
the state we are looking for.

Without the global identity  $d_{-+}=d_{+-}$,
eq.(2.10) would yield two conditions, one for $d_{-+}$ and one for $d_{+-}$.
These reproduce only the detailed balance solution.
The stationary growth solution with $w=1$  is the result of the 
non-local property that there are just as many 
valley bottoms as hill tops.
Every hill top (valley bottom) could have been created during $t \to t+1/N$
with probability $p$ ($q$) and every valley bottom (hill top) 
can be destroyed with the same probability.
Since $d_{-+}=d_{+-}$ the right hand side of eq.(2.11) vanishes
when the coefficients $c(K)$ are independent of $\K$.

\vfill
\eject

\centerline {\bf 3.  THE STATIONARY STATE IN THE RSOS MODEL}
\bigskip

The RSOS growth model describes the growth of simple cubic 
surfaces in which only mono-atomic steps are allowed.
Nearest neighbour columns can differ by only $dh =S^z_n=0,\pm1$.
The master equation is more complex than eq.(2.3) for the BCSOS model,
since more types of local configurations are possible.
Each has its own transition probabilities.
After choosing one of the columns at random,
one particle can be evaporated or deposited at $x=n+\half$ with
a probability that depends on the height differences
$S^z_n$ and $S^z_{n+1}$.
The time evolution operator is again of the form  eq.(2.3)
with 
$$
\eqalignno{
\Ham = \sum_{n=1,N}\{
~[~&p_h~(1-S^+_nS^-_{n+1})~+~q_v(1-S^-_nS^+_{n+1})~]~\delta(0)_n\delta(0)_{n+1} \cr
+~&q_h~(1-S^-_nS^+_{n+1})~\delta(+)_n\delta(-)_{n+1}\cr
+~&p_v~(1-S^+_nS^-_{n+1})~\delta(-)_n\delta(+)_{n+1}\cr
+~&p_s~(1-S^+_nS^-_{n+1})~[\delta(0)_n\delta(+)_{n+1}+\delta(-)_n\delta(0)_{n+1}]\cr
+~&q_s~(1-S^-_nS^+_{n+1})~[\delta(+)_n\delta(0)_{n+1}+\delta(0)_n\delta(-)_{n+1}]~\}
\hskip 0.75 truein
& (3.1)
}
$$
and the following definitions: 
$\delta (0) = (1+S^z)(1-S^z)$,
$\delta (+) = {1\over2} S^z(S^z+1)$, and 
$\delta (-) = {1\over2} S^z(S^z-1)$.
The  raising and lowering operators $S^+$ and $S^-$ 
are normalized such that 
$S^z= S^+S^- - S^-S^+$ and $S^+S^- + S^-S^+= 2 - (S^z)^2$.
This time evolution operator has 6 parameters.
They have the following interpretations:
$p_h$ is the adsorption  probability and 
$q_v$    the evaporation probability  when the surface is locally flat;
$q_h$ is the evaporation probability at   a local hill top;
$p_v$ is the adsorption  probability into a local valley bottom;
$p_s$ is the adsorption  probability and  
$q_s$    the evaporation probability at a step.

$\Ham$ is not Hermitian, but resembles the Hamiltonian of
a  quantum spin-1 chain.
$$
\eqalignno{
\Ham_D (n,n+1) = 
~ c_s               
+&(s_m-2c_s)                (S^z_n)^2
-\quart a_s                  S^z_n S^z_{n+1}
+(c_s-s_m+\quart a_s)       (S^z_n S^z_{n+1})^2\cr
+ &(\half s_d-\quart h_g)    S^z_n S^z_{n+1}(S^z_{n+1}-S^z_{n})
&(3.2a)
}
$$
the diagonal part, and 
$$
\Ham_{OD}(n,n+1) =~- \Flip_+ S^+_nS^-_{n+1} - \Flip_-S^-_nS^+_{n+1}
\eqno(3.2b)
$$
the off-diagonal  part, with
$$
\eqalignno{
\Flip^+ = &(p_h-p_s) (S^z_nS^z_{n+1})^2 
+(p_v-p_s)[1-(S^z_n)^2][1-(S^z_{n+1})^2]  
+p_s
\cr 
\Flip^- = &(q_v-q_s)(S^z_nS^z_{n+1})^2  
+(q_h-q_s) [1-(S^z_n)^2][1-(S^z_{n+1})^2]
+ q_s.
&(3.2c)
}
$$
In $\Ham_D$ we introduced a second notation for the 6 parameters: 
$c_s=p_h+q_v$ is the step creation probability;
$a_s=p_v+q_h$ is the step annihilation probability;
$s_m=p_s+q_s$ is the step mobility;
$f_g=p_h-q_v$ is the growth probability at flat surface areas;
$h_g=p_v-q_h$ is the growth probability at hill tops and valleys; and
$s_d=p_s-q_s$ is the growth probability at steps (step drift).

The phase diagram is only 5 dimensional
because rescaling all 6 parameters by a common factor redefines the
unit of time.
Some familiar models are contained in this master equation as special points
or lines [29,30].
For example, 
the Kim-Kosterlitz model [29] corresponds to $q_v=q_h=q_s=0$ (no evaporation)
and $p_v=p_h=p_s$.

Recall (see section 2)
that the stationary state of the BCSOS model, eq.(2.4), is trivial.
It is the completely disordered state $|D\ket$, because
the adjoint of its time evolution operator it also stochastic.
It describes the time-evolution of the same surface
in the particle-hole transformed representation.
The adjoint of eq.(3.1)
is only stochastic in a 3D subspace of the phase diagram.
The diagonal parts of $\Trans^\dagger$ and  $\Trans$ are identical,
but the coefficients in the off-diagonal part switch position,
$$
\eqalignno{
&\Trans(c_s,a_s,s_m;f_g,h_g,s_d)^\dagger = \cr
&\Trans_D(c_s,a_s,s_m;f_g,h_g,s_d) +
\Trans_{OD}(a_s,c_s,s_m;-h_g,-f_g,-s_d)
&(3.3)}
$$
The parameter space in which $\Trans^\dagger$ preserves probability
is larger than the 2D self-adjoint subspace 
in which the surface does not grow
($p_h=q_h$, $p_v=q_v$, and $p_s=q_s$).
$\Trans^\dagger$ is stochastic whenever the 
diagonal part $\Trans_{D}$ in eq.(3.3), 
is invariant under the transformation 
$c_s \leftrightarrow a_s$
$h_g \leftrightarrow -f_g$, and $s_d \to -s_d$
(the way the parameters switch in the off-diagonal part).
This is true for
$$
c_s=a_s, \quad\quad h_g - 2 s_d = -f_g + 2s_d.
\eqno(3.4)
$$
Inside this 3D subspace the adjoint of the time evolution operator
is stochastic  and describes surfaces in which the probabilities 
are interchanged as
$$
\Trans(c_s,a_s,s_m;f_g,h_g,s_d)^\dagger =
\Trans(a_s,c_s,s_m;-h_g,-f_g,-s_d)
\eqno(3.5)
$$
In the BCSOS model, $\Trans^\dagger$ is identical to 
the particle-hole transformed dynamics.
This is not true anymore in the RSOS model;
the particle-hole transformed time evolution operator
$$
\widehat \OP_{PH}~
\Trans(c_s,a_s,s_m;f_g,h_g,s_d) =
\Trans(c_s,a_s,s_m;-f_g,-h_g,-s_d)
\eqno(3.6)
$$
is different from the one in eq.(3.5).

Next, we generalize this stationary state to a 4D subspace.
The BCSOS model master equation describes
a lattice gas of non-interacting charged particles in an electric field.
The RSOS master equation is a particle-hole  generalization of this. 
The $S^z_n=\pm1$ states represent sites occupied by
particles with electric charge $\pm 1$.
$S^z_n=0$ states represent empty sites.
The energy of such a lattice gas is of the form
$$
\Energy = \sum_n [ \mu (S^z_n)^2+ \EF n~S^z_n].
\eqno(3.7)
$$  
The chemical potential $\mu$ 
is needed, because the dynamic rule does not conserve the 
total number of particles
(the number of steps in the surface).
The chemical potential for the total charge can be omitted because 
electric charge is conserved.
We are looking for a stationary state of the form 
$c(\{S^z_n\}) = c(\N _s, \K)$
in which the coefficients only depend on the number of particles, $\N_s$ and
the total electrostatic energy, $\K= \sum_n n S^z_n$,
$$
c(\{S^z_n\}) \sim \exp[- \mu \N_s - \EF\K]= z^{\half N_s} w^\K.
\eqno(3.8)
$$
Let us try the conventional detailed balance approach for 
Monte Carlo simulations.
Require that the action of each individual 
$\Ham_{n,n+1}$ leave  $|\Psi\ket$ invariant.
This yields the conditions 
$$
\eqalignno{
{q_h\over p_h} = & {c(..,0,0,..) \over c(..,+,-,..)}   \cr
{p_v\over q_v} = & {c(..,0,0,..) \over c(..,-,+,..)}   \cr
{p_s\over q_s} = & {c(..,+,0,..) \over c(..,0,+,..)} 
               =   {c(..,0,-,..) \over c(..,-,0,..)} 
&(3.9)
}
$$
(using the same type of notation as in eq.(2.5)).
They are satisfied by the stationary state of the form eq.(3.8) 
when
$$
\eqalignno{
p_h/q_h ~=~ &\exp[ (\EF - 2\mu)/k_BT]\cr
p_v/q_v ~=~ &\exp[ (\EF + 2\mu)/k_BT]\cr
p_s/q_s ~=~ &\exp[  \EF        /k_BT]
&(3.10)
}
$$
This defines a 4D subspace inside the 5D phase diagram, since 
each set of values $\EF$ and $\mu$ leaves two ratios 
unspecified, $p_h/p_v$ and $p_h/p_s$.

Eq.(3.9) is not the stationary state we are interested in.
It applies to the wrong type of boundary condition.
Let's apply the same non-local approach as in section 2.
Assume that the stationary state is of the form eq.(3.8)
and express the time evolution in terms of 
the 9 nearest neighbour step-step densities $d_{00}$, $d_{+-}$,..., $d_{++}$.
The equations of motion read:
$$
\eqalignno{
{\der c(\N_s,\K) \over \der t} = 
& 
- [(p_h+q_v) d_{00}  +p_v d_{-+} + q_h d_{+-}]~c(\N_s,\K) \cr 
& 
+ p_h d_{+-}~c(\N_s-2, \K+1) + q_v d_{-+}~c(\N_s-2, \K-1) \cr 
&
+ q_h d_{00}~c(\N_s+2, \K-1) + p_v d_{00}~c(\N_s+2, \K+1) \cr
&
+ q_s (d_{0+} + d_{-0})~c(\N_s, \K-1) +  p_s (d_{+0}+d_{0-})~c(\N_s, \K+1) \cr
&
- [p_s (d_{0+} + d_{-0}) + q_s (d_{+0} + d_{0-}) ]~c(\N_s,\K)
&(3.11)
}
$$
This is the generalization of eq.(2.11).
The stationary state is of the form  eq.(3.8),
if the right hand side vanishes for all $d_{i,j}$.
Naively this yields 5 equations with 5 unknowns. 
However, the $d_{i,j}$ in eq.(3.11) are not independent.
The analogues of eq.(2.10) are 
$$
\eqalignno{
d_+ =& d_{++} + d_{+0} + d_{+-} = d_{++} + d_{0+} + d_{-+} \cr
d_- =& d_{--} + d_{0-} + d_{+-} = d_{--} + d_{-0} + d_{-+} 
&(3.12)
}
$$
with $d_+$ and $d_-$ the density of up- and down-steps,
and $d_0+d_++d_-=1$.
Adding and subtracting these two equations yields two identities:
$$
\eqalignno{
2d_{+-} + d_{+0} + d_{0-} = & 2 d_{-+} + d_{0+} + d_{-0} \cr
d_{+0} - d_{0-} = &            d_{0+} - d_{-0} 
&(3.13)
}
$$
After employing the first one, only 
$d_{00}$, $d_{+-}$, $d_{-+}$, and $d_{+0}+d_{0-}$  
remain in eq.(3.11).
Eq.(3.8) is the correct stationary state
if the following  4 conditions are satisfied:
$$
\eqalignno{
(p_h+q_v) ~=~ & q_h z     w^{-1} +  p_v z w \cr
q_h       ~=~ & p_h z^{-1}w      - 2(p_s-q_s w^{-1}) \cr
p_v       ~=~ & q_v (zw)^{-1}    + 2(p_s-q_s w^{-1}) \cr
0 ~=~& (p_s-q_s w^{-1}) (1-w) 
&(3.14) 
}
$$
The last equation indicates that there are two solutions:
one with $q_s/p_s =w$ and one with $w=1$.
The first solution, with $w = q_s/p_s$, reproduces the 
atmospheric law type charge profile (3.9).
The second solution, with $w=1$, applies to periodic boundary conditions 
and is the one we are looking for.
For $w=1$, the last condition in (3.14) is satisfied without
any constraints on $p_s$ and $q_s$. 
Only 2 of the 3 remaining conditions are independent.
One sets the step density and the other specifies 1 condition between 
the 5 growth parameters.
The stationary state is of the form 
$$
c(\{S^z_n\})= c(\N_s) \sim \exp[- \mu \N_s] = (\sqrt z)^{\N_s} 
\eqno(3.15a)
$$
inside  the 4 dimensional subspace defined as:
$$
\eqalignno{
z~=~&{p_h + q_v \over q_h+p_v} \cr
0=~ & z(p_v-q_h) + (p_h-q_v) - 4 z (p_s-q_s)
&(3.15b)  
}
$$
the latter is equivalent to 
$$
\eqalignno{
z =~&c_s/a_s \cr 
4 s_d z =~&z h_g +f_g
&(3.15c)  
}
$$
Both solutions of $w$ are valid in a subspace of co-dimension 1 of the 
5 dimensional parameter space, but these subspaces do not coincide, 
nor do the equations for the step-fugacity $z$.

In the stationary state (3.15) the step density is equal to
$$
s = \bra D| (S^z_n)^2|\Psi\ket = {2 \sqrt z \over 1+2 \sqrt z}   
\eqno(3.16)
$$
for zero surface  tilt, $v= \bra D|S^z_n |S\ket =0$. 
Notice that for $z=1$ we recover the disordered state solution, eq.(3.4).

The stationary growth rate, $r_g$, is equal to 
$$
\eqalignno{ 
r_g 
= & ~\bra D| (\Flip_+ S^+_nS^-_{n+1} - \Flip_-S^-_nS^+_{n+1}) |\Psi\ket \cr
= & f_g (1-s)^2 + \quart h_g s^2 + s_d s(1-s)
&(3.17)
}
$$
which simplifies with eq.(3.15c) and eq.(3.16) into
$$
r_g =  s_d s
\eqno(3.18)
$$
The height-height correlation function diverges linearly
because the spins are uncorrelated:
$$
\bra D| (h_{n+\half}- h_{n+m+\half})^2 |\Psi\ket =
\bra D| \sum_{n<i\leq n+m} (S^z_{i})^2 |\Psi\ket = m s .  
\eqno(3.19)
$$
This confirms that the surface roughness critical exponent 
is equal to  $\alpha=\half$.

It is amazing that the stationary state is
this simple in such a large fraction of the phase diagram.
The step-step correlation function, eq.(1.4), has a sharp cut-off, 
to such an extent that steps are completely disordered.
Outside this 4D subspace the stationary state becomes more complex.
Eq.(1.4) does not have a sharp cutoff anymore.
We checked this numerically.
Attempts to extend the above derivation 
to a form in which the $c(\{ S_n  \})$ are functions only
of the nearest neighbour step-step correlations are doomed.
The number of equations increases rapidly 
and they can not be satisfied inside the 5D parameter space.

\vfill
\eject

\centerline {\bf 4. MEAN FIELD THEORY AND LANGEVIN EQUATIONS}
\bigskip

In this section we apply mean field theory to the BCSOS and RSOS model.
The mean field equations of motion for the order parameters 
reproduce the deterministic part of the KPZ equation.
This is similar to equilibrium critical phenomena,
where mean field theory for the Ising model
reproduces only the non-fluctuating  part of $\phi^4$-theory.
The object is to better understand the structure of the 5D RSOS model phase diagram;
in particular, to relate qualitatively the parameters of the 
RSOS model to those in the KPZ equation.

Consider the BCSOS Master equation, eq.(2.4).
The equation of motion
$$
{\der v\over \der t} = v_n(t+1)-v_n(t)= \bra D|[S^z_n,\Trans]|\Psi\ket_t
\eqno(4.1)
$$
for the local slope of the surface
$$
v_n(t) = \bra D|S^z_n|\Psi\ket_t
\eqno(4.2)
$$
does not close because of the presence of 
nearest neighbour step-step correlations on the right hand side.
Define 
$$
d(a,b)_x = \bra D| \delta( S^z_{n}-a) \delta( S^z_{n+1}-b) |\Psi\ket
\eqno(4.3)
$$
with $a,b = \pm 1$ and $x=n+\half$, and  
replace the slope variable by the 
surface height using the relation $v_n= h_{x+1}-h_x$.
Eq.(4.1) yields,
$$
{\der h\over \der t} = p ~d(+,-)_x - q~d(-,+)_x
\eqno(4.4)
$$  
This is the top of a BBGKY  hierarchy of equations of motion.
In lowest order approximation all step-step correlation are neglected. 
The $d(a,b)_x$ are assumed to factorize. 
The equation of motion for the average local slope $v_n(t)$ then closes.
$$
{\der  \over \der t }~h_{x} ~= 
{1\over 4} p (1-v_n)(1+v_{n+1}) - {1\over 4} q (1+v_n)(1-v_{n+1}). 
\eqno(4.5)
$$
A faster route to exactly the same equation is to apply the
mean field approximation directly to the master equation. 
The adsorption and evaporation transition rates
at site $x$ are assumed to depend 
on the expectation values of the step density $v_{n}$ 
instead of their actual values $S^z_{n}$. 
Eq.(4.5) is identical to the deterministic part of the KPZ equation,
eq.(1.1), in the spatial continuum limit  
$$
{\der h \over \der t} ~= ~
{1\over 4}  (p-q)
+{1\over 4} (p+q)   {\der^2 h \over \der x^2} 
-{1\over 4} (p-q)   ({\der h   \over \der x  })^2
\eqno(4.6)
$$
$\lambda$ is negative and reduces the growth rate at slopes.
This reflects that in the BCSOS model sloped surface sections, 
with $S_n=S_{n+1}$, are inactive.

Apply the same type of approximation to the RSOS model.
The first observation is that the 
RSOS model contains two point-like expectation values instead of one:
the local step density, $s_n$, and the local slope of the surface, $v_n$,
$$
\eqalignno{
&s_n(t) = \bra D|(S^z_n)^2|\Psi\ket_t\cr
&v_n(t) = \bra D|S^z_n|\Psi\ket_t.
& (4.7)
}
$$
Mean field theory involves two order parameters instead of one.
The equations of motion for these two order parameters, 
$$
\eqalignno{
&{\der s\over \der t} = s_i(t+1)-s_i(t)= \bra
D|[(S^z_i)^2,\Trans]|\Psi\ket_t &(4.8.a)\cr
&{\der v\over \der t} = v_i(t+1)-v_i(t)=
\bra D|S^z_i,\Trans]|\Psi\ket_t &(4.8.b)\cr
}
$$
are again the top of a BBGKY type hierarchy 
of coupled differential equations: 
$$
\eqalignno{
{\der s_{x+{1\over2}} \over \der t} =
&(p_h+q_v) [d(0,0)_{x} + d(0,0)_{x+1}] \cr
&- p_v     [d(-,+)_{x} + d(-,+)_{x+1}]
-  q_h     [d(+,-)_{x} + d(+,-)_{x+1}] \cr
&+ p_s     [d(-,0)_{x} - d(0,+)_{x} - d(-,0)_{x+1} + d(0,+)_{x+1}] \cr
&+ q_s     [d(+,0)_{x} - d(0,-)_{x} - d(+,0)_{x+1} + d(0,-)_{x+1}] 
& (4.9.a)\cr
&\cr 
{\der h_{x} \over \der t} =
&(p_h-q_v) d(0,0)_{x} + p_v d(-,+)_{x} - q_h d(+,-)_{x} \cr
&+    p_s [d(-,0)_{x} + d(0,+)_{x}]
-     q_s [d(+,0)_{x} + d(0,-)_{x}]
& (4.9.b)
}
$$
In  mean field theory the nearest neighbour step-step
correlations factorize.
The resulting equations are not very transparent. 
Therefore, we first consider the special case where the surface
is  spatially  uniform, such that $s=s_n$ and $h=h_x$ do not depend
on position.
The mean field approximation for eq.(4.9)  then reads
$$
\eqalignno{
{\der s \over \der t} =
&2 c_s  (1-s)^2 - \half a_s   (s-v)(s+v) & (4.10.a)\cr
&\cr 
{\der h_{n} \over \der t} =
&f_g (1-s)^2  + \quart h_g (s^2-v^2)  + s_d (1-s)s & (4.10.b)
}
$$
The equation for the step density contains a ``mass" term, which
defines a characteristic time $\tau_s$. 
At zero net tilt, $v=0$,
the step density reaches its stationary value 
$$
s_0= [1+\half \sqrt{a_s \over c_s})]^{-1}
\eqno(4.11.a)
$$ 
after a characteristic time
$$
\tau_s^{-1} = 2\sqrt{a_s c_s}
\eqno(4.11.b)
$$ 
At time scales larger than $\tau_s$ 
the surface grows at a constant average rate
$$
r_0= f_g (1-s_0)^2 + \quart h_g s_0^2+ s_d(1-s_0)s_0 
\eqno(4.11.c)
$$
This means that, although there are two order parameters, only one of them,
the local slope $v_n$, fluctuates at time scales larger than $\tau_s$. 
The full mean field equation for the step density reads:
$$
\eqalignno{
{\der s\over \der t} = 
&~~2c_s(1-s)^2
- \half a_s(s^2-v^2)
+  s_d {\der\over\der x}[(1-s)v] 
+ \half  h_g (v {\der s \over \der x} - s{\der v \over \der x})
\hskip 0.5 truein \cr
&+ (\half s_m-c_s)    {\der^2 s \over \der x^2} 
+      \quart a_s  v  {\der^2 v \over \der x^2} 
+ (c_s-\quart a_s) s  {\der^2 s \over \der x^2} 
~,
& (4.12) \cr
}
$$
in which all derivatives are discrete.
Define $\Delta$ as the deviation from the stationary value, $s=s_0+\Delta$.
At time scales larger $\tau_s$ 
the step density does not behave as an independent dynamic variable.
$\Delta$ follows local fluctuations in the slope of the surface
such that the right hand side of eq.(4.12) remains equal to zero:
$$
m_s \Delta  = \delta \left({\der h \over \der x}\right)^2 
+\epsilon {\der^2 h \over \der x^2} ...
\eqno(4.13)
$$
with
$\delta    = \half  a_s$ and 
$\epsilon  = s_d(1-s_0)-\half h_g s_0$.
The full equation of motion for $h_x$ is complex.
It contains many terms,
involving the step density, derivatives of the surface slope,
and combinations of these and their derivatives.
At time scales larger than $\tau_s$, the step density
is not an independent variable,
and can be eliminated using eq.(4.13).
The equation for $dh/dt$ then 
reduces to the deterministic part of the KPZ equation, eq.(1.1), with 
$$
\eqalignno{
\lambda     \simeq  &~ \half  a_s\tau_s [  h_gs_0 - 4f_g(1-s_0) +2s_d(1-2s_0)] - \half h_g, \cr
\nu         \simeq  &~ \half s_m(1-s_0) +\quart a_s s_0.
& (4.14)}
$$
and many higher order terms as well. 
The above two derivations 
illustrate that the BCSOS and RSOS growth models belong 
to the KPZ universality class. 
That is hardly a surprise however.
The significance is the identication of an additional order parameter
and of the time scale $\tau_s$. 
At short time scales the step density
behaves as an independent dynamic variable
and the KPZ equation description is incomplete.
Fortunately $\tau_s$ is typically very short, 
of the order of only a few time steps.

\vfill
\eject

\centerline {\bf 5. SKEWNESS IN THE RSOS MODEL}
\bigskip

The phase diagram of the RSOS model seems too large to
visualize, but the results of the previous sections 
put some order into it.
The uncorrelated stationary state (in the subspace defined by eq.(3.15))
and the mean field theory define a set of characteristic
densities and time scales.
The six parameters in eq.(3.2) fall into two groups:
$c_s$, $a_s$, and $s_m$, are equilibration type parameters;
$f_g$ $h_g$ and $s_d$, are growth type parameters. 

The ratio between the step creation and 
annihilation probabilities, $z=c_s/a_s$, controls the steps density, $s$.
Their product $\tau_s^{-1}= 2 \sqrt{a_sc_s}$ controls the time
scale at which the step density equilibrates.
The step mobility $s_m$ is a suitable unit of time. 
(We set $s_m=1$.)
KPZ type scaling is realized at time scales larger than $\tau_s$ 
and length scales larger than the inverse of the step density.
This leaves us with an effective 3D phase diagram,
characterized by the growth parameters,
$f_g$, $h_g$, and $s_d$.
The linear combination 
$$
r_g= f_g (1-s^2) + \quart h_gs^2 + s_d s(1-s),
\eqno(5.1)
$$
with $s$ defined in eq.(3.15),
controls the growth rate.
The linear combination
$$
u_{sk}=zh_g+f_g-4z s_d
\eqno(5.2)
$$
controls the skewness of the stationary state, see eq.(3.15).
In the zero skewness plane, $u_{sk}=0$,
$r_g$ is  exactly equal to the growth rate 
and $s$ is exactly equal to the step density.
The third independent combination 
of $f_g$ $h_g$ and $s_d$  controls the 
strength of the non-linear term of the KPZ equation.
Eq.(4.14) is an approximation for $\lambda$.

The phase diagram of the RSOS model is effectively only three dimensional,
but is still larger than the one for the KPZ equation, eq.(1.1). 
The RSOS model allows us to control skewness.
Stationary skewness can turned on in the KPZ equation as well.
The stationary state of eq.(1.1) is known exactly in 1D. 
It is a Gaussian distribution without any skewness [5,10,27].
This is an accident.
Additional operators in the equation should introduce stationary state skewness.
The following $\lambda_2$ type term is an example of this 
$$
{d h \over d t} = r_g +\nu \nabla^2 h + \half \lambda (\nabla h)^2 + 
\lambda_2 (\nabla^2 h)^2
+\eta.
\eqno(5.3)
$$
We performed some (qualitative) Monte Carlo runs at $\lambda_2\neq0$  [34]. 
Those confirm that the stationary state is skewed.
The BCSOS growth model and the $u_{sk}=0$ RSOS model
represent special cuts through the extended KPZ equation, 
where parameters such as $\lambda_2$ take special values
such that the stationary state lacks skewness.
On a qualitative level the role of $\lambda_2$ 
can be understood as follows:
In a flat surface, the growth rate in the RSOS model is 
determined by $f_g$, and in the KPZ equation by $r_g$.
In a sloped surface, the growth rate in the RSOS model is
determined by $f_g(1-s_0) + s_ds_0$,
and in the KPZ equation by $r_g+\lambda$.
In a hilly surface without skewness with densely packed steps,
the growth rate in the RSOS model is determined by $h_g$, 
and in the KPZ equation, eq.(5.3), by $r_g+\lambda_2$.

We hoped that mean field theory (section 4) would provide a meaningful 
estimate of the $u_{sk}=0$ space in the generalized KPZ equation.
Unfortunately, too many operators seem to be involved.
For example, the analysis of section 4 gives a non-zero
value for $\lambda_2$ in the BCSOS model. 
Apparently that value for $\lambda_2$ is compensated by significant 
contributions of other operators such that stationary state 
skewness remains absent.
 
The reverse route might be more promising.
The major difference between eq.(5.3) and microscopic models like 
the BCSOS and RSOS model
is the manner in which the noise couples to the local surface structure.
In the Langevin equation they are ``additive".
The probablity distribution obeys a Fokker-Planck equation [5,35].
For example, in discretized time it is allowed to visualize the
time evolution as determinsitic between $t\to t+\half$ and
purely stochastic (ballistic deposition like) between $t+\half \to t+1$.
$$
\eqalignno{
h(t+\half) =& h(t) + r_g +\nu \nabla^2 h + \half \lambda (\nabla h)^2 +
\lambda_2 (\nabla^2 h)^2 \cr 
h(t+1)     =& h(t+\half) + \eta. &
(5.4)
}
$$
In the BCSOS and RSOS model the noise couples much more directly to the 
local surface configuration.
Correlation effects between steps  are more pronounced.
 
The KPZ fixed point must lie inside the non-skewness subspace.
As a rule, fixed points lie in subspaces 
where the short distance properties mimic as much as possible 
the proper long distance scaling properties.
For example, the fixed point of the Ising model on a square lattice lies
in the subspace where the next nearest neighbour interactions are
of the same order of magnitude as the nearest neighbour interactions,
such that the correlation functions are rotational invariant 
not only at large distances, but also at short distances.
The cubic anisotropy is an irrelevant operator and needs to be 
turned-off at the fixed point.
  
The following numerical results demonstrate that 
stationary state skewness is indeed irrelevant in KPZ type growth.
We determine the largest eigenvalues and eigenvectors
of the time evolution operator for system sizes $4 \leq L\leq 14$. 
These values are exact, and allow
a detailed finite size scaling (FSS) analysis.
This method is identical to conventional transfer matrix
FSS calculations in equilibrium phase transitions,
and works equally well for master equations  [13].
We applied this method to several points in the phase diagram,
but we present only our results for the 
Kim-Kosterlitz (KK) model [29].
This is a typical point outside the $u_{sk}=0$ subspace.

On a local level skewness implies an imbalance between sharp hill tops
and valleys bottoms.
The quantity 
$$
\rho_{sk} = {d(+,-)-d(-,+) \over d(+,-)-d(-,+) } 
\eqno(5.5)
$$
measures the density difference between sharp hill-tops 
and sharp valley-bottoms. 
$\rho_{sk}$ converges to a non-zero value at the KK point, as shown in Fig.2.
This illustrates the presence of skewness in the stationary state 
at short distances.
The long distance probe for skewness is the third moment, $W_3$, of the
height distribution function, see eq.(1.2). 

The scaling behaviour of $W_3$ tells us whether 
stationary state skewness is present or absent at the KPZ fixed point.
It should diverge as $W_3 \sim L^{3 \alpha}$ if present.
Otherwise, we must incorporate into eq.(1.3) the crossover scaling in the 
skewness direction $u_{sk}$,
$$
W_n(L,t,u_{sk}) = b^{n\alpha} W_n(b^{-1}L, b^{-z}t, b^{y_{sk}} u_{sk}).
\eqno(5.6)
$$
and expand this in small $u_{sk}$,
$$
W_n(L,\infty, u_{sk}) \simeq L^{n\alpha} \left[ W(1,\infty,0) +  
L^{y_{sk}} u_{sk} W^\prime(1,\infty,0)+....\right]
\eqno(5.7)
$$
In the absence of skewness 
the amplitude $W_3(1,\infty,0)$ of the leading term is equal to zero
and the third moment scales as $W_3 \simeq A L^x$
with $x=3\alpha+y_{sk}$.
Fig.3 demonstrates that stationary state skewness is indeed 
absent at the KPZ fixed point.
The exponent $x=0.4 \pm 0.1$ suggests that the
skewness crossover exponent is equal to $y_{sk}=-1$
(the nearest integer).
The amplitude $A$ of the third moment, see  Fig.~4, must be proportional to $u_{sk}$.
The local measure for the skewness, $\rho_{sk}$, in Fig.~2 
should be approximately proportional to $u_{sk}$ as well.
We calculated the ratio $A/\rho_{sk}$ at several points 
in the phase diagram, and find it indeed to be almost a constant. 

In conclusion, 
in this paper we study the phase diagram of the RSOS growth model in
1+1 dimensions. Its phase diagram contains a 4D subspace in which the 
stationary state is completely uncorrelated. Familiar models, such as the
Kim-Kosterlitz model, lie outside this subspace. Their stationary states
contain additional features, like skewness. 
Stationary state skewness diverges with system size, 
but such that the scaling properties are still described by the 
KPZ fixed point at zero skewness. The skewness crossover
exponent is equal to $y_{sk}\simeq-1$.

This research is supported by NSF grant DMR-9205125.
\vfill
\eject

\centerline {\bf REFERENCES}
\bigskip
\item{ 1. } 
M.~Kardar, G.~Parisi, and Y-C.~Zhang,Phys.~Rev.~Lett. {\bf 56}, 889 (1986).
\item{ 2. } 
J.~Krug and H.~Spohn in {\it Solids Far from Equilibrium: Growth,
Morphology and Defects}, ed. C. Godr\`eche (Cambridge University Press,
Cambridge, 1991).
\item{ 3. } 
P.~Meakin, Physics Reports {\bf 235}, 189 (1993).
\item{ 4. } 
J.~Krug in {\it Scale Invariance, Interfaces, and Non\--Equilibrium
Dynamics},ed. A.~McKane, M.~Droz, J.~Vannimenus, and D.~Wolf 
(Plenum, NY, 1995).
\item{ 5. } 
T.J.~Halpin-Healy and Y.C.~Zhang, Physics Reports {\bf 254}, 215 (1995).
\item{ 6. }
D.~Dhar, Phase Transitions {\bf 9}, 51 (1987).
\item{ 7. } 
L-H.~Gwa and H.~Spohn, Phys.~Rev.~Lett.~{\bf 68}, 725 (1992);
and, Phys.~Rev.~A {\bf 46}, 844 (1992).
\item{ 8. }
D.~Kim, Phys.~Rev.~E {\bf 52}, 3512 (1995).
\item{ 9. }
M.R.~Evans and B.~Derrida, Acta Physica Slovaca {\bf 44}, 331 (1994).
\item{10. }
D.~Forster, D.R.~Nelson, and M.J.~Stephen, Phys.~Rev.~A {\bf 16}, 732 (1977).
\item{11. }
M.~Kardar and Y-C.~Zhang, Phys.~Rev.~Lett.~{\bf 58}, 2087 (1987).
\item{12. }
L.~Golubovi\'c and Z-G.~Wang, Phys.~Rev.~E {\bf 49}, 2567 (1994).
\item{13. } 
J.~Neergaard and M.~den~Nijs, Phys.~Rev.~Lett. {\bf 74}, 730 (1995).
\item{14. }
B.~Derrida and M.R.~Evans, in {\it Nonequilibrium
Statistical Mechanics in One Dimension}, ed. V.~Privman
(Cambridge University Press, 1996).
\item{15. }
B.~Derrida, M.R.~Evans, and D.~Mukamel, J.Phys.~A {\bf 26}, 4911 (1993).
\item{16. }
G.~Sch\"{u}tz and E.~Domany, J.Stat.Phys.~{\bf 72}, 277 (1993).
\item{17. }
J.D.~Noh and D.~Kim, Phys.~Rev.~E {\bf 53}, 3225 (1996).
\item{18. }
E.~Medina, T.~Hwa, M.~Kardar, and Y-C.~Zhang, Phys.~Rev.~A {\bf 39}, 3053 (1989).
\item{19. }
B.~Nienhuis, in {\it Phase Transitions and Critical Phenomena, vol 11},
ed. C.~Domb and J.~Lebowitz (Academic, London, 1987).
\item{20. }
M.~den~Nijs, in {\it Phase Transitions and Critical Phenomena, vol 12},
ed. C.~Domb and J.~Lebowitz (Academic, London, 1988).
\item{21. }
J.L.~Cardy, in {\it Phase Transitions and Critical Phenomena, vol 11},
ed. C.~Domb and J.~Lebowitz (Academic, London, 1987).
\item{22. }
M.~Henkel, J.~Stat.~Phys.~{\bf 75}, 1023 (1994).
\item{23. }
Per Fr\"ojdh and Marcel~den~Nijs, unpublished.
\item{24. }
J.~Krug, P.~Meakin, and T.~Halpin-Healy, Phys.~Rev.~A {\bf 45}, 638 (1992).
\item{25. }
J.~Krug and H.~Spohn, Phys.Rev.~A {\bf 38} 4271 (1988).
\item{26. }
D.E.~Wolff, J.Phys.~A {\bf 20}, 1251 (1987).
\item{27. }
U.~Deker and F.~Haake, Phys.~Rev.~A {\bf 12}, 1629 (1975).
\item{28. }
P.~Meakin, P.~Ramanlal, L.~Sander, R.C.~Ball, Phys.~Rev.A {\bf 34}, 5091 (1986).
\item{29. }
J.~M.~Kim and J.~M.~Kosterlitz, Phys.~Rev.~Lett.~{\bf 62}, 2289 (1989).
\item{30. }
J.G.~Amar and F.~Family, Phys.~Rev.~Lett.~{\bf 64}, 543 (1990). 
\item{31. }
J.~D.~Shore and D.~J.~Bukman, Phys.~Rev.~Lett.~{\bf 72}, 604 (1994).
\item{32. }
H.~van Beijeren and I.~Nolden, in {\it Strucures and dynamics of
Surfaces}, edited by W.~Schommers and P.~von Blanckenhagen  Vol.~2
(Springer, Berlin 1987).
\item{33. }
E.H.~Lieb and F.Y.~Wu,
in {\it Phase Transitions and Critical Phenomena}
eds. C.~Domb and M.S.~Green (Academic, London, 1972).
\item{34. }
John~Neergaard, Ph.D.~Thesis, University of Washington (1996).
\item{35. }
K.~Park and B-N.~Kahng, Phys.~Rev.~E {\bf 51}, 796 (1995)
\vfill
\eject
\centerline{\bf FIGURE CAPTIONS}
\bigskip
\item{ 1. } 
Temporal skewness in the deterministic KPZ equation.
Time evolution of a sinusoidal initial state.
\item{ 2. } 
Finite size scaling of the local skewness order parameter, $\rho_{sk}$,
eq.(5.5), at the Kim-Kosterlitz point.
$\rho_{sk}$ represents the density difference between sharp hill 
tops and sharp valley bottoms. 
\item{ 3. } 
The scaling exponent $x$ of the third moment, $W_3\simeq AN^x$,
at the Kim-Kosterlitz point.
The drawn line represents finite size scaling estimates 
$x(N)$ from the exact values of $W_3(N)$ at
N-1 and N+1 for  $N=5,7,..13$.
The dashed line shows estimates for $x$ 
defined as $x(N)\simeq x+B/N^2$ at successive values of $N$.
\item{ 4. }
Finite size scaling estimates of the amplitude $A$ of the third moment, 
$W_3\simeq AN^x$, at the Kim-Kosterlitz point, assuming that $x=0.4$.
The dashed line represents the same type of extrapolation as
in Fig.3.
\vfill
\eject
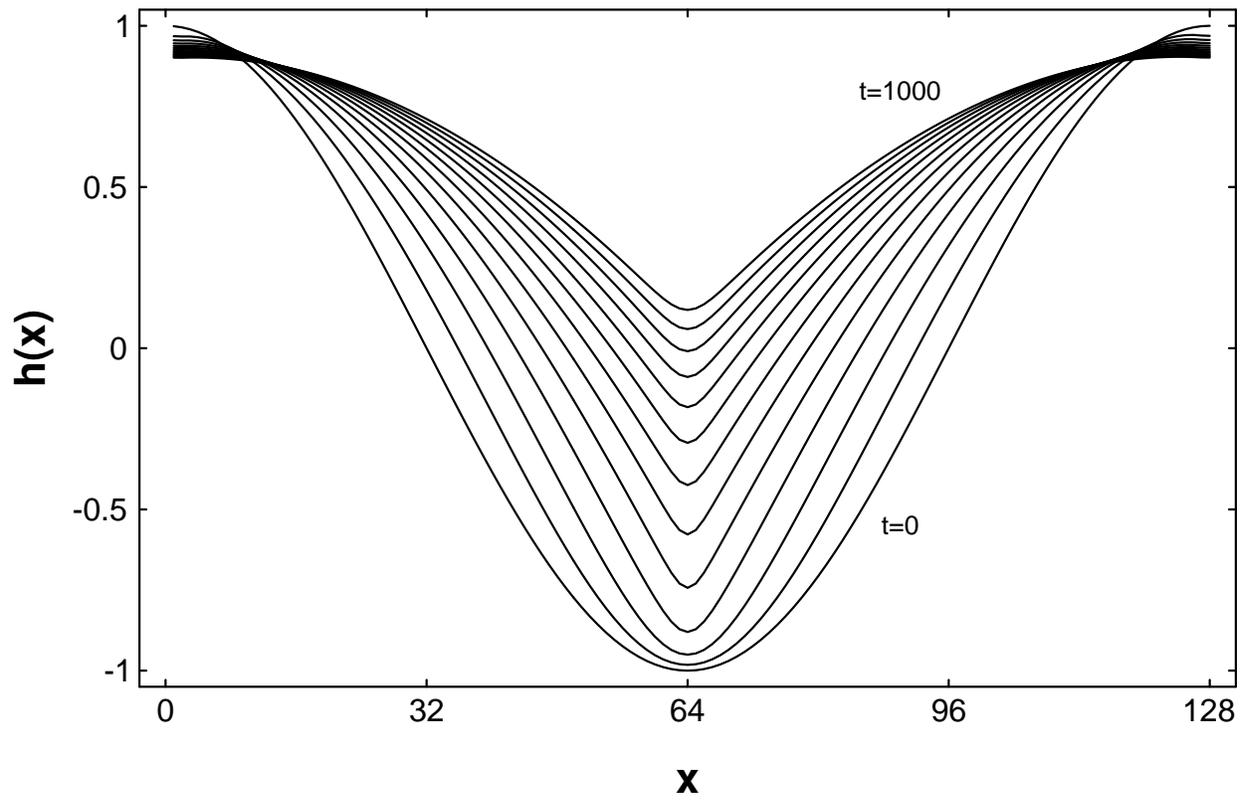
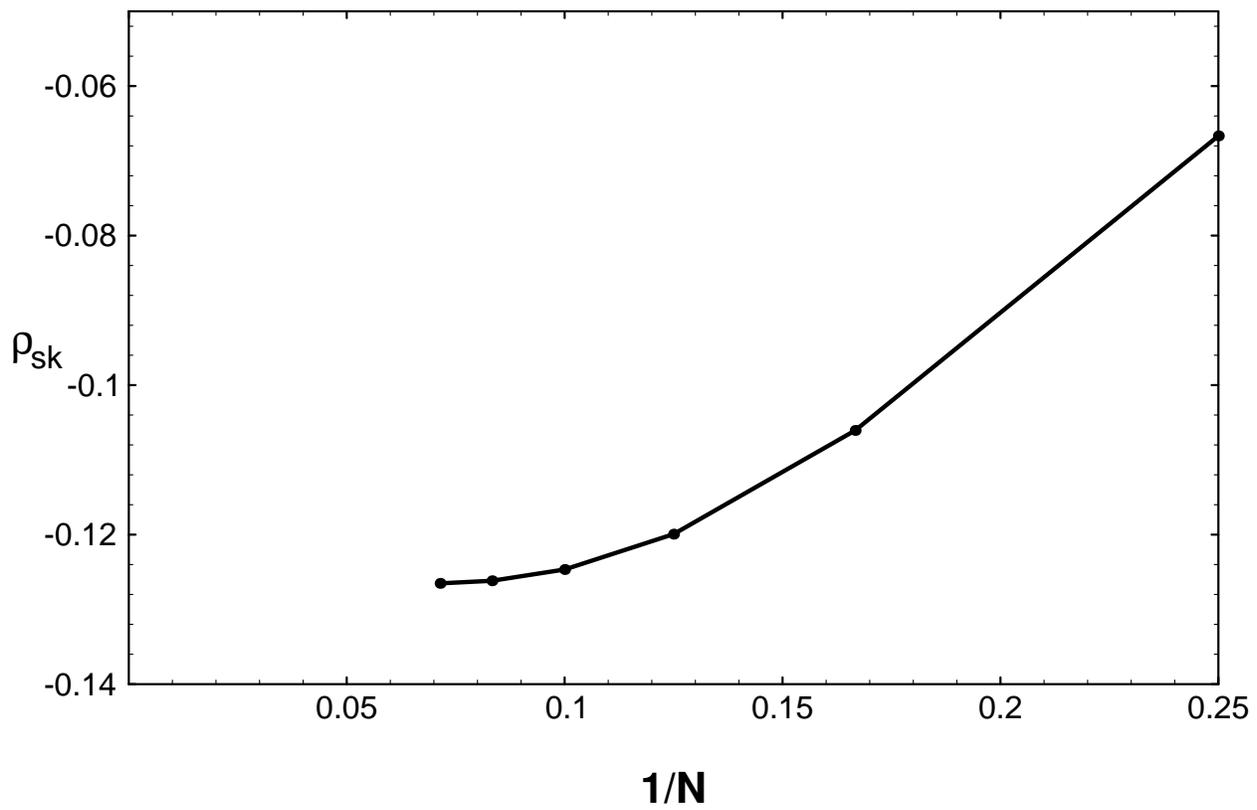
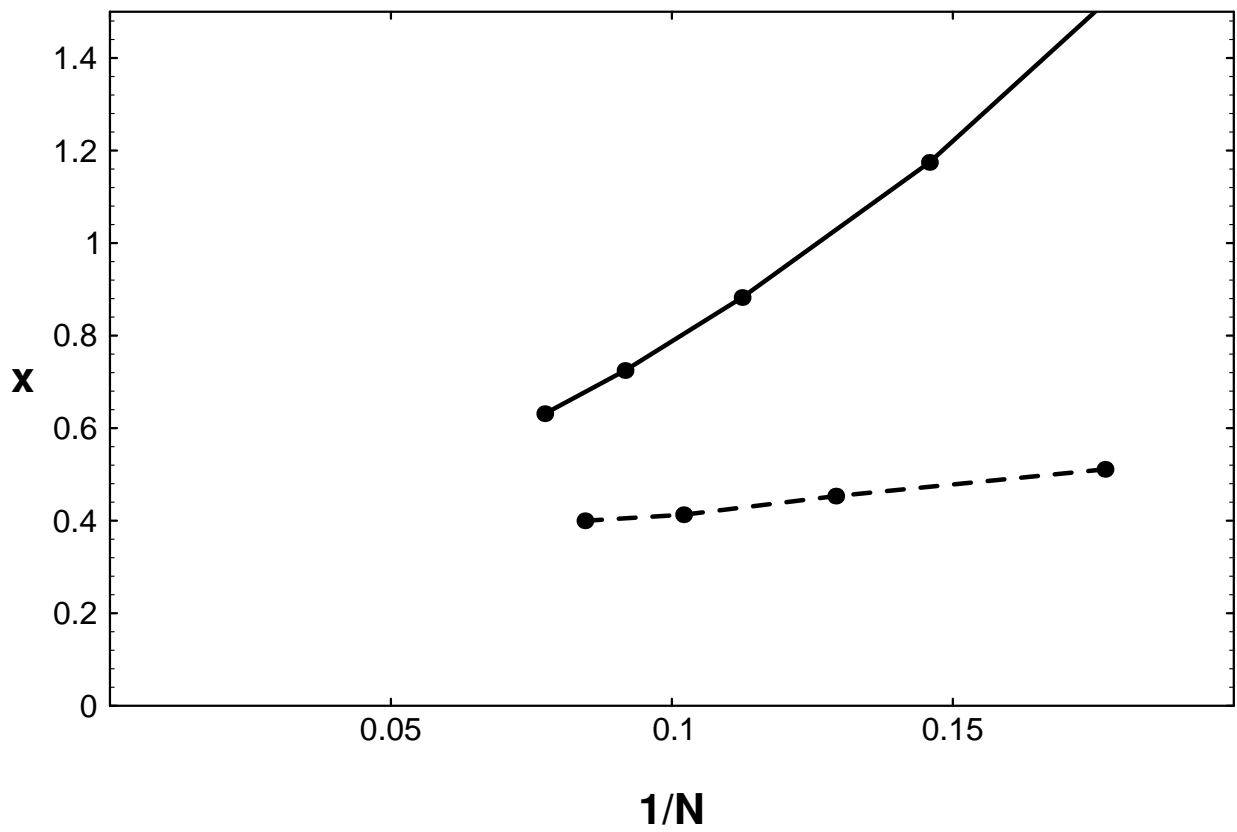
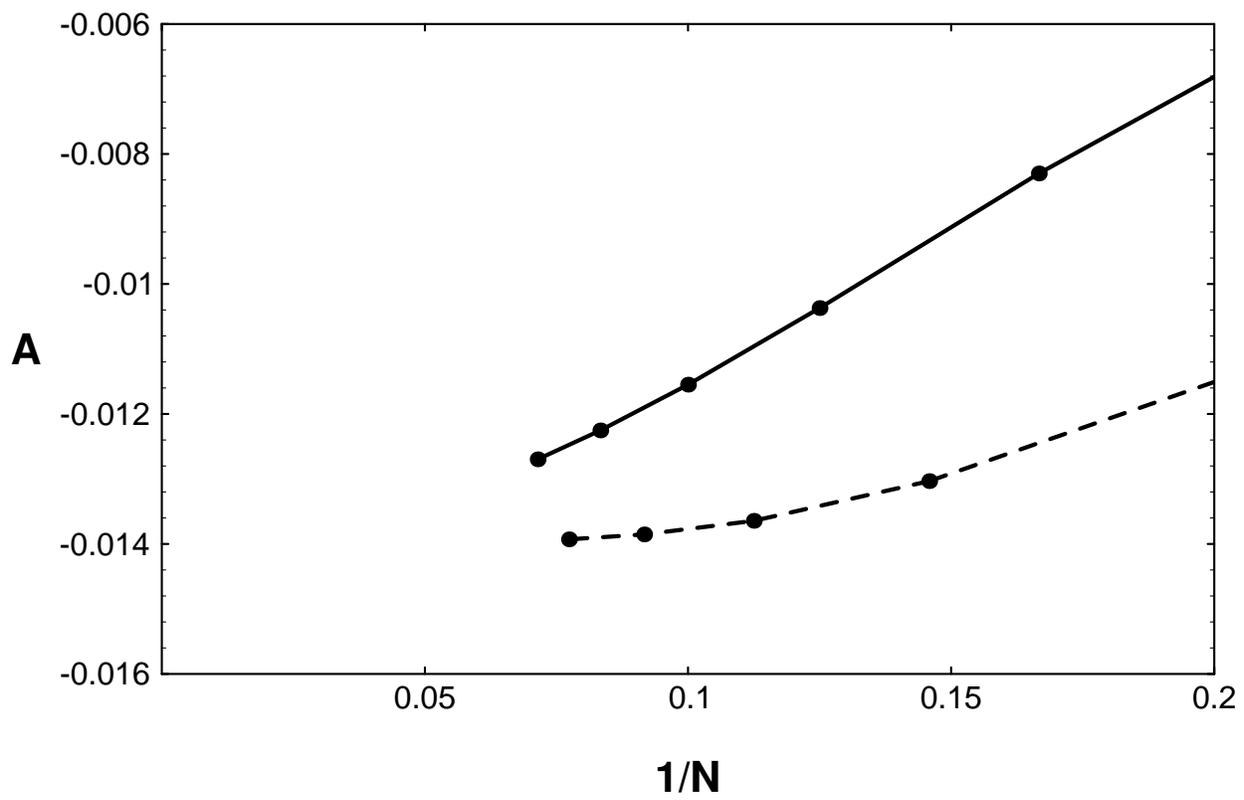
\end